\documentclass[fleqn,usenatbib]{mnras}

\usepackage{newtxtext,newtxmath}
\usepackage{caption}
\usepackage{multicol}
\usepackage{slashbox}
\usepackage{multirow}
\usepackage{subcaption}

\usepackage[T1]{fontenc}

\DeclareRobustCommand{\VAN}[3]{#2}
\let\VANthebibliography\thebibliography
\def\thebibliography{\DeclareRobustCommand{\VAN}[3]{##3}\VANthebibliography}

\usepackage{graphicx}	
\usepackage{amsmath}	
\usepackage{amssymb}	

\date{Accepted 2024 January 19. Received 2024 January 10; in original form 2023 June 29}

\pubyear{2024}

\title[Radio pulsar surface magnetic field]{Effect of variable crustal density on the surface magnetic field of Radio Pulsars}

\author[Sellick \& Ray]
{Kathleen Sellick\thanks{$\!\!$e-mail:kas.sellick@gmail.com } and
Subharthi Ray\thanks{$\!\!$e-mail:rays@ukzn.ac.za } \\
Astrophysics Research Centre (ARC), School of Mathematics, Statistics and Computer Sciences, \\ University of KwaZulu-Natal, Durban 4001, South Africa}
\date{Accepted 2024 January 19. Received 2024 January 10; in original form 2023 June 29}

\pubyear{2024}

\begin{document}
\maketitle
\begin{abstract}
We study the surface magnetic field fluctuations due to radial oscillations as a viable cause for the micro structures of the radio pulsar pulse patterns. The electrical conductivity of matter in the outer layer of the crust of a neutron star (NS) plays a crucial role in the resulting surface magnetic field if we assume that the magnetic field is confined to this layer.  This outer layer has a rapidly varying matter density - that changes the micro-physics of the material affecting the electrical conductivity at every stage of the density change. In this study, the varying electrical conductivity in this rapidly varying density regime of the outer layer of the NS crust - from $\sim 10^{11}~g~cm^{-3}$ to about $10^4~g~cm^{-3}$ - has been used to calculate the surface magnetic field using the induction equation. A finite effect of the strong gravitational field at the NS surface has also been taken into account. The equations have been solved in MATLAB using the \emph{method of lines}. Any minor radial fluctuation due to stellar oscillation, in particular the radial oscillations, leads to a fluctuation of the electrical conductivity in the outer layer of the crust. This leads to fluctuations in the surface magnetic field with a frequency equal to the frequency of the stellar oscillation. We find that not only the variation of the surface magnetic field is substantial, but also it does not remain constant throughout the lifetime of the NS.

\end{abstract}


\section{Introduction}
\label{sec:intro} 

In over five decades since their discovery, the radio pulsars still present us with challenges from theoretical and mathematical understanding of the physical processes going on in that system. The magnetic fields of neutron stars play a pivotal role in all of the observed activities of the stars. The magnetospheric structure of a NS is believed to be made of closed and open field lines, where the closed field lines form a torus around the star and co-rotate with it. This region is charge saturated - with the charges mainly originating from pair creation of electron and positrons due to the strong surface magnetic field. The closed field lines \emph{"open"} up at the light cylinder and the observed emission emerge from this conical zone. Though mostly the emission from the light cylinder is in the form of a wind of plasma, a small fraction of this plasma is converted to a coherent beam of radio waves arising due to the growth of plasma instabilities in the inner magnetosphere, from regions below 10\% of the light cylinder radius (\citet{mitra2017}). This beam is ejected relentlessly, which we finally observe as the pulse profile - also suggests that the pair creation process generating the plasma has to be continuous. The magnetic field of the magnetosphere is the strongest at the NS surface, where the pair creation (and hence the generation of the plasma) is maximum. Since the closed field lines are charge saturated, the surface magnetic field at the polar caps plays a seed role in the subsequent processes for the creation of highly polarised radio beams (and the winds).

There is no conclusive theory to explain the actual origin of the NS magnetic field. It is generally accepted that the strong magnetic field of a NS is due to flux conservation of the original star from which it was born. Said so, there isn't any \emph{a priori} knowledge about the distribution of the magnetic field inside the NS. There are numerous mechanisms which may result in field generation in the crust of NSs, for instance, \citet{bah1983} and \citet{uly1986} explored the idea that thermomagnetic instability may have this result. The primary problem in hosting a super strong magnetic field in the NS core is due to the fact the nuclear matter in the core is highly super-conducting, resulting in repulsion of flux to a lower density regime. Another challenge may come from general relativity (GR), where the vectorial nature of the dipolar magnetic field does not allow interior solutions for a spherically symmetric configuration for realistic equations of state (\citet{ram2001}). Detailed discussions about these problems can be found elsewhere - for our present work, we shall focus on the crustal magnetic field evolution only and accept that the magnetic field vanishes in the deep layers. 

Considering the magnetic field being located in the outer layer of the crust, the behaviour of the surface magnetic field will then depend largely on the transport properties, in particular the electric and the thermal conductivities of the matter in the outer crust and how the field evolves through it. One of the earlier works in connection to the transport properties of dense matter was carried out by \citet{flit1976}. Thereafter notable studies on the electrical and thermal conductivities were done by \citet{yakur1980}, \citet{yakur1980a}, \citet{yakur1980b}, \citet{flit1981}, \citet{itoh1984}, etc., where we find ways of calculating the electrical conductivity due to phonon scattering and impurity scattering. 

The magnetic field of a NS decays over it's entire lifetime. Ohmic decay of dipolar magnetic fields, that are confined to the outer NS crust, was first studied by \citet{sangchan1987} who concluded that the field does not decay exponentially but by less than a factor of order $10^2$ in the Hubble time. \citet{chansang1989} calculated the braking index of neutron stars from the ohmic decay of the magnetic fields, and showed that the surface magnetic fields may fluctuate and even change polarities at short time scales. \citet{ucs1994} derived an analytic solution to the asymptotic behaviour of the dipolar magnetic fields. \citet{urmus1992} studied the ohmic decay of the magnetic field confined to the surface layers of the neutron star crust with consideration for the neutron star cooling. They showed that the initial depth penetrated by the magnetic field is important and has a notable effect on the results. The conductivity of the crust also depends largely on the thermal state of the matter and the cooling process of the star. \citet{urpvan1993} compared the magnetic field decay for two neutron stars consisting of two different core matters and thus two different cooling models, finding that the accelerated cooling model has much less field decay than the standard model. Other notable studies on neutron star cooling processes include the works of \citet{riper1991}, \citet{yak2004}, \citet{apm2008} and \citet{ppp2015}.
 
\citet{batdat1996} studied the ohmic diffusion of the magnetic flux expelled from the core of a neutron star to its crust, and inferred it to be caused from the spindown-induced field evolution. This study was done for multiple NS equations of state for which they took the crustal density profiles from \citet{dattaetal1995}. \citet{dattaetal1995} noted that there are many papers published which considered cooling models for a specific equation of state model and then used a crustal model that correspond to a different equation of state model, which they attribute to a non-availability of detailed numerical crustal density profiles. 

\citet{sen1997} similarly considered the ohmic decay of the magnetic field confined to the crust of the NS and studied for both the flat and curved spacetimes - neglecting the NS cooling. He showed that GR has a significant effect on the results, particularly the lengthening of the decay time of the magnetic field. Later \citet{sen1998} included the effects of the neutron star cooling in both flat and curved spacetimes and obtained results comparable to \citet{urmus1992}, again showing that the GR effects reduce the decay rate substantially. \citet{geppage2000} also performed comparisons of the flat and curved spacetimes and similarly found that the GR effects increased the decay time of the magnetic field. Further studies on ohmic decay and Hall drift in the NS crust have been done by \citet{pongep2007}, \citet{blb2018}, \citet{pmg2009}, \cite{igo2021} and \cite{ponsvig2019}. 

In order to explain radio emission state changes in pulsars, \cite{gbmms2021} proposed the rapid modification of the neutron star surface magnetic field as the primary contributor to the process. According to them, the combined effect of Hall and thermal drift causes oscillations of the local magnetic field structure at the surface of the polar cap of radio pulsars, leading to the rapid  modifications of the surface magnetic field which can even be of the order of a single pulse period.

Borrowing the idea from \cite{gbmms2021}, in this work we have investigated the possibility of generating a variability in the surface magnetic field that can come from the radial fluctuation of the outer layer of the NS crust, if we consider the NS magnetic field is confined to this outer later. The incompressibility of nuclear matter is extremely high, and hence in any oscillatory mode of a NS, the fluctuations at the outer crust will be more pronounced than the nuclear matter core. This fluctuation in the outer crust will lead to a density fluctuation and subsequently a fluctuation in the conductivity of matter, thus affecting the resultant surface magnetic field. Observationally, pulsars exhibit dynamic pulse patterns containing complex microstructures, such as  sub-pulses, which have durations of a few microseconds to nanoseconds within one single pulse period. This observed behaviour should then be directly related to the behaviour of the magnetic field at the NS surface.

In the literature, radial oscillations in neutron stars (including strange stars) for zero temperature EOS have been studied by \cite{chan1977}, \cite{vc1992}, \cite{gll1983}, etc, and for finite temperature EOS by \cite{gon1997}, \cite{gz1999}, \cite{bh1991}, etc. A comprehensive study of radial oscillation using multiple NS EOSs has been done by \cite{koru2001}, where they found that the fundamental mode of such oscillations are of the order of KHz. \cite{mhh1988} studied various oscillation modes of a NS and have also shown that the crustal fluctuation during such oscillations are much more than the fluctuations of the core.  

 In this paper, in section \ref{sec:form} we calculate the magnetic field evolution through the outer layer of the NS -  using the induction equation, and in a general relativistic background. We also show how the electrical conductivity of matter behaves in the outer layer of the crust. A short discussion on the physical nature of the outer crust is made in section \ref{sec:nscrust}. In section \ref{sec:res} we show our numerical findings, where we estimate the surface magnetic fields for two values of surface fluctuations $\sim$ 0.2\% and 0.4\% of the crust width. We notice a sizeable variation of the surface field observed within a short time scale. We finally conclude in section \ref{sec:concl} with some plausible inferences.

\section{Basic Formalism}
\label{sec:form}

In our study we have used the induction equation in a general relativistic background for the evolution of the magnetic field $\mathbf{B}$ through the outer layer of the NS crust. We have also shown how the electrical conductivity will behave in the rapidly varying density regime of the outer crust. 

\subsection{Evolution of the magnetic field in a general relativistic background}

The induction equation in a flat spacetime is given by:

\begin{equation}
\label{eq:induction}
\frac{\partial \mathbf{B}}{\partial t} = - \boldsymbol{\nabla} \times \mathbf{E} 
\end{equation}
where 
\begin{equation}
\mathbf{E} = \frac{c^2}{4\pi \sigma} \boldsymbol{\nabla} \times \mathbf{B}.
\nonumber
\end{equation}

Assuming the magnetic field to be a pure dipole field, the induction equation in a curved spacetime is derived using the covariant form of Maxwell's equations,

\begin{equation}
\label{eq:covMax1}
\frac{1}{\sqrt{-g}} \frac{\partial}{\partial x^\nu} \left( \sqrt{-g}F^{\mu\nu} \right) = - \frac{4\pi}{c} J^\mu
\end{equation}
and

\begin{equation}
\label{eq:covMax2}
\frac{\partial F^{\mu\nu}}{\partial x^\lambda} + \frac{\partial F^{\nu\lambda}}{\partial x^\mu} + \frac{\partial F^{\lambda\mu}}{\partial x^\nu} = 0,
\end{equation}
with the generalised Ohm's law
\begin{equation}
\label{eq:genOhm}
J^\mu = \sigma g^{\mu\nu} F_{\nu\lambda} u^\lambda
\end{equation}
where $F_{\mu\nu}$ are the components of the electromagnetic field tensor, $J^\mu$ are the components of the four-current density, $u^\mu$ are the components of the four velocity of the fluid, $g_{\mu\nu}$ are the components of the spacetime metric describing the background geometry, and $g = det  \lvert g_{\mu\nu} \rvert $. In our notation Latin indices represent the 3-dim spatial coordinates and Greek indices represent the 4-dim space-time coordinates.

If we take into account a stationary gravitational field, then using equations (\ref{eq:covMax1}), (\ref{eq:covMax2}), and (\ref{eq:genOhm}) and taking $u^i = dx^i/ds = 0$ we can derive the corresponding induction equation in a curved spacetime as:

\begin{eqnarray}
\label{eq:indcurve}
\nonumber
\frac{\partial F_{kj}}{c\partial t} =  \frac{c}{4\pi}\left\{ \frac{\partial}{\partial x^k}\left[ \frac{1}{\sqrt{-g}}\frac{1}{\sigma u^0}g_{ij}\frac{\partial}{\partial x^l}\left( \sqrt{-g}F^{il} \right) \right] \right. \\
 \left. - \frac{\partial}{\partial x^j}\left[ \frac{1}{\sqrt{-g}}\frac{1}{\sigma u^0}g_{ik}\frac{\partial}{\partial x^l}\left( \sqrt{-g}F^{il} \right) \right] \right\}.
\end{eqnarray}

In the outer crust of a NS, the magnetic field strength will be relatively weaker compared to the gravitational field strength present there, so that the spacetime curvature is not affected by the magnetic field. Hence it is reasonable to assume the magnetic field configurations of the NS crust in a background Schwarzschild geometry - described by the metric:

\begin{equation}
\label{eq:metric}
ds^2 = \left( 1 - \frac{2m}{r} \right) c^2dt^2 - \left( 1-\frac{2m}{r} \right)^{-1} dr^2 - r^2 \left( d\theta^2 + \sin^2{\theta} d\phi^2 \right)
\end{equation}
where $m = MG/c^2$ with $M$ being the total gravitational mass of the core.  The crust consists of less than a few percent ($<3\%$) of the total gravitational mass and hence $M$ can be approximated to the total mass of the star. Hence the self-gravitation of the crust is negligible when compared to the gravitational field due to the core. 

The non-zero components of the orthonormal tetrad $ e^\gamma_{(\alpha)}$ of the local Lorentz frame for the Schwarzschild geometry is:

\begin{eqnarray}
\nonumber
e^t_{(t)}   = \left( 1 - \frac{2m}{r} \right)^{-\frac{1}{2}}; & ~~
e^r_{(r)}  = \left( 1 - \frac{2m}{r} \right)^{\frac{1}{2}};~~
\nonumber
e^\theta_{(\theta)}  = \frac{1}{r}; ~~~ \\
\nonumber
e^\phi_{(\phi)}  = \frac{1}{r\sin{\theta}}.~~~~~~~~~~&
\end{eqnarray}

The components of the electromagnetic field tensors for the local Lorentz frame ($F_{(\alpha\beta)}$) and that of the curved spacetimes ($F_{\gamma\delta}$ ) are then connected to each other with the above tetrads as:

\begin{equation}
\label{eq:FLorentz}
F_{\left( \alpha\beta \right)} = e^\gamma_{\left( \alpha \right)} e^\delta_{\left( \beta \right)} F_{\gamma\delta}.
\end{equation}

Making use of the metric in equation (\ref{eq:metric}), the induction equation in curved space (equation (\ref{eq:indcurve})) reduces to:
\begin{eqnarray}
\label{eq:redindcurve}
\nonumber
\frac{\partial F_{kj}}{c\partial t} =  \frac{c}{4\pi}\left\{ \frac{\partial}{\partial x^k}\left[ \frac{1}{r^2 \sin{\theta}}\frac{1}{\sigma u^0}g_{ij}\frac{\partial}{\partial x^l}\left( r^2 \sin{\theta}F^{il} \right) \right] \right. \\
 \left. - \frac{\partial}{\partial x^j}\left[ \frac{1}{r^2 \sin{\theta}}\frac{1}{\sigma u^0}g_{ik}\frac{\partial}{\partial x^l}\left( r^2 \sin{\theta}F^{il} \right) \right] \right\}.
\end{eqnarray}

Since we are considering the decay of a dipolar magnetic field with axial symmetry, the vector potential $\mathbf{A}$ can be written, in spherical polar coordinates, as $(0,0,A_\phi)$ where $A_\phi = A(r,\theta,t)$. The metric then gives $u^0$ as:
\begin{equation}
\label{eq:u0}
u^0 = \left( 1 - \frac{2m}{r} \right)^{- \frac{1}{2}}.
\end{equation}

Using this in equation (\ref{eq:redindcurve}) we obtain:  
\begin{eqnarray}
\label{eq:redindcurve_rphi}
\nonumber
\frac{\partial F_{r\phi}}{\partial t}  = \frac{c^2}{4\pi} \frac{\partial}{\partial r}  \left[ \frac{1}{\sigma} \left( 1 - \frac{2m}{r} \right)^{ \frac{1}{2}}  \sin{\theta} \left\{ \frac{\partial}{\partial r} \left[ \left( 1 - \frac{2m}{r} \right) \frac{F_{r\phi}}{\sin{\theta}} \right] \right. \right.\\
\left. \left. + \frac{\partial}{\partial \theta} \left( \frac{F_{\theta\phi}}{r^2 \sin{\theta}} \right) \right\} \right]
\end{eqnarray}

and

\begin{eqnarray}
\label{eq:redindcurve_thetaphi}
\nonumber
\frac{\partial F_{\theta\phi}}{\partial t}  = \frac{c^2}{4\pi} \frac{\partial}{\partial \theta}  \left[ \frac{1}{\sigma} \left( 1 - \frac{2m}{r} \right)^{ \frac{1}{2}} \sin{\theta} \left\{ \frac{\partial}{\partial r} \left[ \left( 1 - \frac{2m}{r} \right) \frac{F_{r\phi}}{\sin{\theta}} \right] \right.\right.\\
\left.\left. + \frac{\partial}{\partial \theta} \left( \frac{F_{\theta\phi}}{r^2 \sin{\theta}} \right) \right\} \right].
\end{eqnarray}

We then use the definition

\begin{equation}
\label{eq:FAdef}
F_{\alpha\beta} = A_{\beta,\alpha} - A_{\alpha,\beta}
\end{equation}
to find the induction equation in Schwarzschild geometry in terms of the vector potential as:

\begin{eqnarray}
\label{eq:indSchwar}
\nonumber
\frac{\partial A_\phi}{\partial t} = \frac{c^2}{4\pi\sigma}  \left( 1 - \frac{2m}{r} \right)^{\frac{1}{2}} \sin{\theta} \left\{ \frac{\partial}{\partial r} \left[ \left( 1 - \frac{2m}{r} \right) \frac{1}{\sin{\theta}} \frac{\partial A_\phi}{\partial r} \right] \right. \\
 \left. + \frac{\partial}{\partial \theta} \left( \frac{1}{r^2\sin{\theta}} \frac{\partial A_\phi}{\partial \theta} \right) \right\}.
\end{eqnarray}

\noindent We choose the vector potential for a flat spacetime as:

\begin{equation}
\label{eq:Aflat}
A_\phi = \frac{f(r,t)}{r} \sin{\theta}
\end{equation}

\noindent and that for a curved spacetime as:

\begin{equation}
\label{eq:Acurved}
A_\phi = - g(r,t) \sin^2{\theta}.
\end{equation}

Using this with equations (\ref{eq:induction}) and (\ref{eq:indSchwar}) we obtain equations for the flat and curved spacetimes respectively as: 

\begin{equation}
\label{eq:flatdiff}
\frac{c^2}{4\pi R^2\sigma(x)}\left[\frac{\partial^2f\left( x,t \right)}{\partial x^2} - \frac{2}{x^2}f\left( x,t \right)\right] =\frac{\partial f\left( x,t \right)}{\partial t}
\end{equation}

and

\begin{eqnarray}
\label{eq:curveddiff}
\nonumber
\frac{c^2}{4\pi R^2\sigma(x)}\left( 1 - \frac{y}{x} \right)^{\frac{1}{2}} \left[ \left( 1 - \frac{y}{x} \right) \frac{\partial^2 g\left( x,t \right)}{\partial x^2} + \frac{y}{x^2} \frac{\partial g\left( x,t \right)}{\partial x}\right.\\
\left.  - \frac{2}{x^2} g\left( x,t \right) \right] 
= \frac{\partial g\left( x,t \right)}{\partial t} 
\end{eqnarray}

\noindent where $R$ is the radius of the star, $\sigma(x)$ is the variable conductivity of the outer crust, $x = r/R$ and $y = 2m/R$. In order to solve the above two differential equations, we need to impose initial conditions and boundary conditions. If the initial potential (at $t = 0$) for a flat spacetime is given by: 

\begin{equation}
\label{eq:initialpotflat}
A_\phi\left( r,\theta,0 \right) = A_\phi\left( r,\theta \right) = \frac{f\left( r,0 \right)}{r} \sin{\theta} = \frac{f\left( r \right)}{r} \sin{\theta}
\end{equation}

\noindent then for a curved spacetime we have:

\begin{equation}
\label{eq:initialpotcurved}
A_\phi\left( r,\theta,0 \right) = A_\phi\left( r,\theta \right) = - \frac{3rf\left( r \right)}{8m^3} s\left( r \right) \sin^2{\theta} 
\end{equation}

\noindent where $s(r)$ is the general relativistic correction factor. Thus, if we assume the initial value for a flat spacetime to be $f(r,t) = f(r)$, then for a curved spacetime (\citet{wassha1983})

\begin{equation}
\label{eq:g0}
g\left( r,0 \right) = g(r) = \frac{3rf(r)}{8m^3} \left[ r^2\ln{\left( 1 - \frac{2m}{r} \right)} + 2mr + 2m^2 \right].
\end{equation}

The boundary conditions, which are valid for all times $t$, for both the flat and curved spacetimes are given by (\citet{batdat1996})

\begin{subequations}
\begin{align}
\label{eq:boundcond1}
R\frac{\partial g}{\partial r} + g & = 0 \text{ at } r = R \\
\label{eq:boundcond2}
g & = 0 \text{ at crust bottom }.
\end{align}
\end{subequations}

\subsection{Electrical conductivity of the outer crust}

Before we can determine the nature of the electrical conductivity, we need to establish the dependence of the density $\rho$ with the depth from the surface $d=R-r$. In the outer layers of the crust, there is a drastic variation in the density. Thus one needs to have a consistent model matching the density with the depth. \citet{dattaetal1995} computed this variation for various equations of state of nuclear matter. A simple but elegant model was given by \citet{uryak1979} which closely resembles the results of \citet{dattaetal1995}. We have adopted this model here in order to find our solutions. The analytical formula describing this model are as follows:

\begin{equation}
\label{eq:chi}
\chi = \sqrt{z(z+2)} = \left( \frac{\rho}{\mu_e 10^6 \text{g }\text{cm}^{-3}} \right)^{\frac{1}{3}},
\end{equation}

\begin{equation}
\label{eq:z}
z = \frac{d}{H_R},
\end{equation}

\begin{equation}
\label{eq:HR}
H_R = \frac{m_e R^{2}}{m m_p \mu_e}\left( 1-\frac{2m}{R} \right)^{1/2},
\end{equation}
where $m_e$ and $m_p$ are the electron and proton masses respectively, $\mu_e = A/Z$ and $A$ and $Z$ are the atomic number and charge number respectively. For simplicity we assume that only one species of ions is present.

From (\ref{eq:chi}) one can obtain the density in terms of the radius as 

\begin{equation}
\label{eq:rhor}
\rho(r) = \left[ \left(R-r\right)\left(R-r+2H_R\right) \right]^{3/2} \frac{1}{H_R^3} \mu_e 10^6 ~~{\rm g ~cm^{-3}}.
\end{equation}

The electrical conductivity has two components. The first is due to the phonon scattering or oscillatory scattering of the electrons - that happens more in the denser region, where electrons are more free. The second is due to the impurity scattering. The conductivity due to phonon scattering $\sigma_{ph}$ is given by \citet{yakur1980} from which we can write $\sigma_{ph}$ as

\begin{equation}
\label{eq:sigmaph}
\sigma_{ph} = \frac{1.57 \times 10^{23} \chi^4}{T_6(2 + \chi^2)} \frac{\sqrt{0.017 + \delta^2}}{13\delta} ~~\text{ s}^{-1}.
\end{equation}

\noindent Here $T_6 = T/10^6 \text{K}$ and $\delta = 0.45(T/T_D)$. $T_D$ is the Debye temperature given by

\begin{equation}
\label{eq:Debye}
T_D = 0.45 \frac{\hbar}{k_B}\left( \frac{4\pi Z^2 e^2 n_i}{A m_p} \right)^{\frac{1}{2}} = 2.4 \times 10^6 \left( \frac{2}{\mu_e} \right)^{\frac{1}{2}} \chi^{\frac{3}{2}} \text{ K}
\end{equation}
where $k_B$ is the Boltzmann constant and $n_i$ is the ion number density.

The conductivity for impurity scattering $\sigma_{imp}$, which dominates at lower temperatures is given by:

\begin{equation}
\label{eq:sigmaimp}
\sigma_{imp} = \frac{8.53 \times 10^{21} \chi^3}{\Lambda_{imp} \left( 1 + \chi^2 \right)} \frac{Z}{Q} ~~\text{ s}^{-1}.
\end{equation}

\noindent Here $\Lambda_{imp}$ is the Coulomb logarithm which is $\Lambda_{imp} \cong 2$ for $\rho\geq 10^5 ~~\text{ g cm}^{-3}$ and $Q$ is the impurity parameter:

\begin{equation}
\label{eq:imppar}
Q = \frac{1}{n} \sum_i n_i (Z - Z_i)^2
\end{equation}
where $n$ and $Z$ are the number density and electric charge respectively of background ions in the crust lattice without impurity and $n_i$ and $Z_i$ are the density and charge of the $i\text{th}$ impurity species. The summation is extended over all species of impurities. This impurity parameter is independent of the depth and time. Due to lack of an exact understanding of the extent of the impurity content in the crust and the types of impurity species present there, an approximate range of $Q$ varying from 0.1 to 0.001 has been assumed in most of the earlier studies found in literature.

The net electrical conductivity within the crust is then given by: 

\begin{equation}
\label{eq:sigma}
\sigma = \left( \frac{1}{\sigma_{ph}} + \frac{1}{\sigma_{imp}} \right)^{-1}.
\end{equation}

\noindent Making substitutions for $\sigma_{ph}$ and $\sigma_{imp}$, we can write the electrical conductivity in terms of the density $\rho$ as:

\begin{eqnarray}
\label{eq:sigden}
\nonumber
\sigma(\rho) = \left[ 8.28 \times 10^{-29} T \left( \frac{\mu_e 10^6}{\rho} \right)^{4/3} \left[ 2 + \left( \frac{\rho}{\mu_e 10^6} \right)^{2/3} \right] \right. \\ \nonumber
 \times \left[ 9.67 \times 10^{11} \frac{1}{T^2\mu_e^2} \frac{\rho}{10^6} + 1 \right]^{-1/2} \\
 \left. + 1.17 \times 10^{-22}\Lambda_{imp} \frac{Q}{Z} \frac{\mu_e 10^6}{\rho} \left[ 1 + \left( \frac{\rho}{\mu_e 10^6} \right)^{2/3} \right] \right]^{-1}~~{\rm s^{-1}}.
\end{eqnarray}

In Fig.~\ref{fig:sig_den_diffQ}, we showed the variation of the conductivity of matter ($\sigma$) against the matter density ($\rho$) in the outer layer of the crust for three values of the impurity parameter $Q$ = 0.1, 0.01 and 0.001. We see that at higher density ($\sim 10^{11} {\rm g ~cm}^{-3}$) the conductivity changes by about one order of magnitude for a similar change in the impurity parameter - which gradually converges and unifies from a density of $10^7~{\rm g ~ cm}^{-3}$ and lower. The temperature has been kept constant at $T=10^{5.2}$K.

\noindent
\vskip 0.2cm
\begin{minipage}{\linewidth}
\makebox[\linewidth]{
  \includegraphics[keepaspectratio=true,scale=0.33, angle=-90]{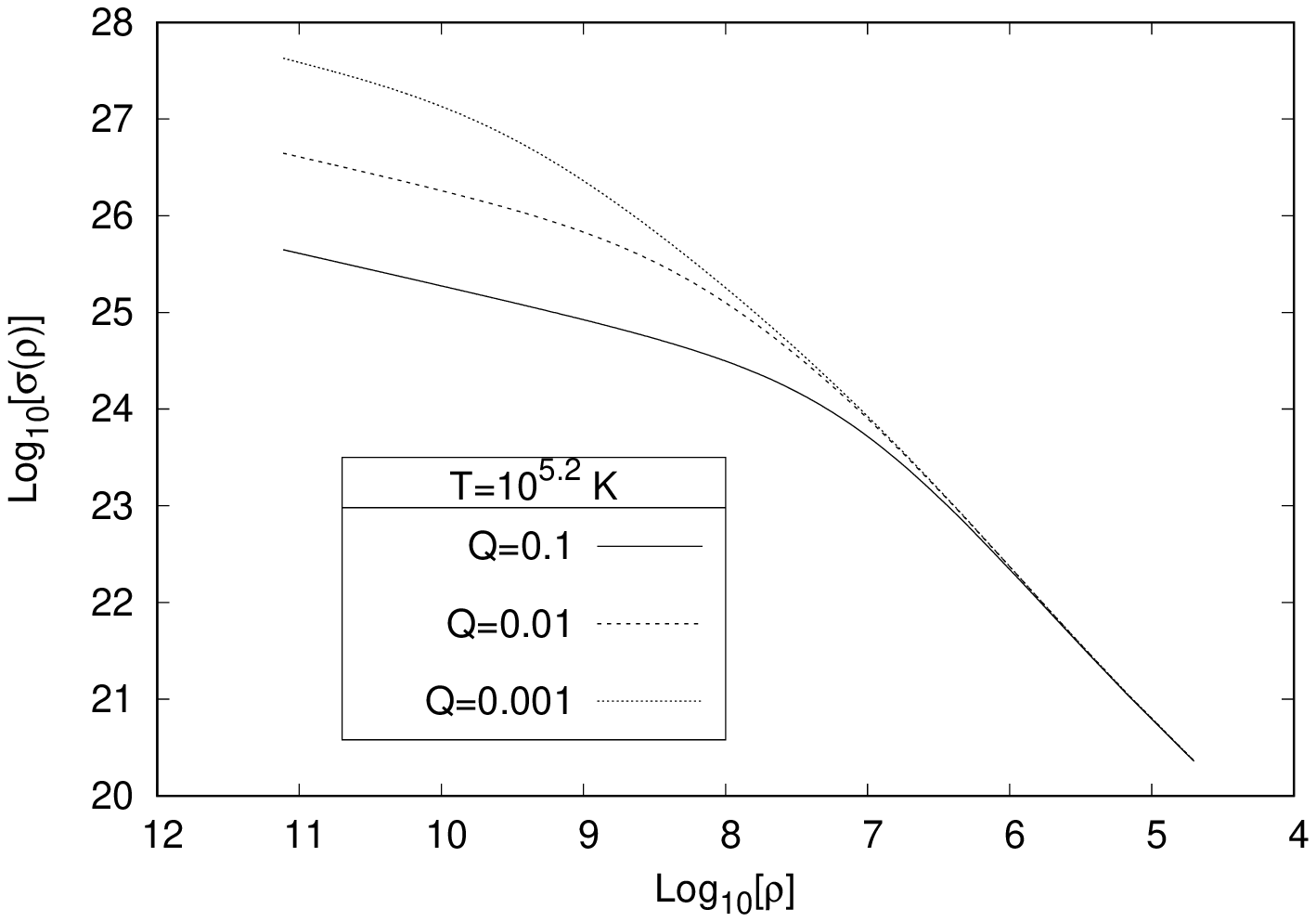}}
\captionof{figure}{Variation of the conductivity ($\sigma$) with the density of matter ($\rho$) near the stellar surface for different values of the $Q$ parameter, and for a temperature $T=10^{5.2}$K.}\label{fig:sig_den_diffQ}
\end{minipage}
\vskip .3cm

Fig.~\ref{fig:sig_den_diffT} shows the variation of the conductivity ($\sigma$) against the matter density ($\rho$) in the outer layer of the crust for three different values of temperature $T=10^{5}$K, $10^{5.2}$K \& $10^{5.5}$K, for a fixed value of the impurity parameter $Q=0.001$. We note that in the lower density regime, $\sigma$ varies substantially with slight variation of temperature.

\noindent%
\vskip 0.2cm
\begin{minipage}{\linewidth}
\makebox[\linewidth]{
  \includegraphics[keepaspectratio=true,scale=0.33, angle=-90]{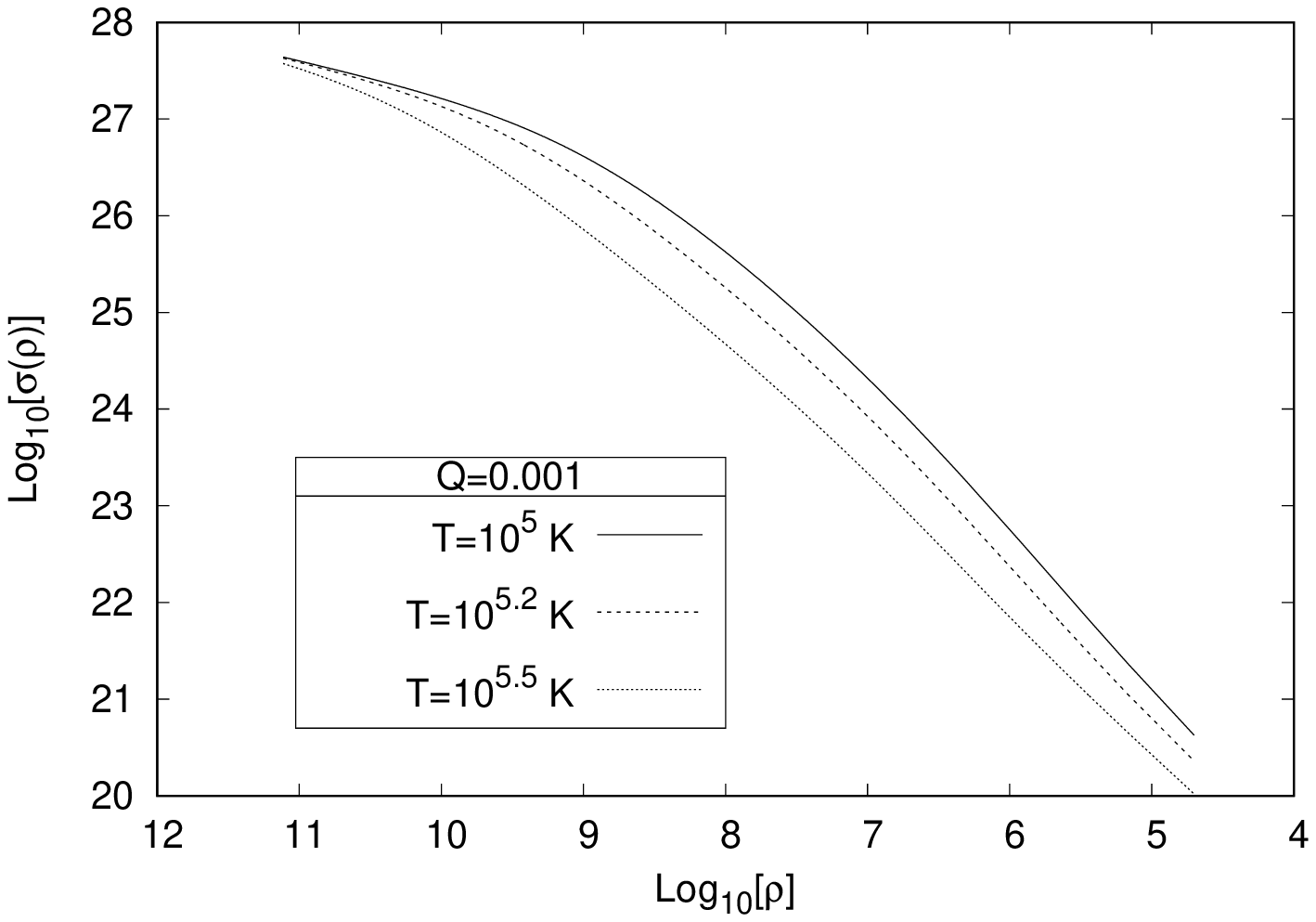}}
\captionof{figure}{Variation of the conductivity ($\sigma$) with the density of matter ($\rho$) near the stellar surface for three different values of the temperature, and for $Q=0.001$.}\label{fig:sig_den_diffT}  
\end{minipage}
\vskip .2cm

\noindent We also note that the electrical conductivity changes by  nearly 7 to 8 orders of magnitude in the outer layer of the crust that is spanning only about 500 metres.

\section{The outer layer of the NS crust}\label{sec:nscrust}
Typically  neutron stars have a nuclear matter core in the density range of $\rho \sim 10^{14} g ~cm^{-3}$. The crusts are believed to be about the outer 10\% of the stellar radius, where the density changes from $\sim 10^{14} g ~cm^{-3}$ to zero at the surface. In such a wide density variation the nature of the particles inside the crust also change drastically (\cite{bp75}). Near the surface, above $\rho \sim 10^{4} g ~cm^{-3}$ the electron kinetic energy rises with the density. With increase in the matter density, there is an increase in the electrical conductivity of the crust till it reaches the \emph{neutron drip} ($\rho \sim 10^{11} g ~cm^{-3}$). Up to this point the matter consists of nuclei of the normal metal in a sea of electrons - providing the charge neutrality. Beyond the \emph{neutron drip}, the matter becomes energetically favourable for neutron continuum states and the nuclei becomes immersed in a mixture of neutron and electron gas. The bottom end of the neutron star crust is the normal nuclear matter density, where there are mostly neutrons and a smaller fraction of protons and electrons.

The last few hundred metres, which is defined as the outer crust, has a very rapid variation in the density of the matter. It is this region which is important for our present study. The rapidly varying electrical conductivity in this rapidly varying density region of the outer crust is also affected by the temperature of the matter. A neutron star during its birth, attains the temperature of the supernova process, which is of the order of MeV ($\sim 10^{9-10}$K). However this initial temperature quickly cools down by a few orders of magnitude by the neutrino emission in a direct URCA process \citep{ykgh2001} from the nuclear matter core and by the neutrino bremsstrahlung from electron-nucleus collision in the crust \citep{ch2008}.  In a non-linear relativistic mean field model of NS EOS with protons ($p$), neutrons ($n$) and hyperons ($\Xi, ~\Sigma, ~\Lambda$), with particle interaction via the exchange of $\sigma$, $\rho$ and $\omega$ mesons \citep{glen1989}, it has also been shown that a relatively high proton fraction triggers a direct URCA process \citep{lpph1991}, thus inducing an enhanced cooling. Modified URCA process involving presence of additional nucleon in the direct URCA reactions, have also been modeled \citep{YaLe1995} as an alternative mechanism for a NS cooling. \cite{rodrigo2011} and \cite{rodrigo2012} have incorporated rotational deformation to estimate the thermal evolution of a NS to show that from about $10^2$ years the polar and equatorial temperatures fall to about $10^{5.2}$K after which the temperature remains relatively constant throughout the lifetime of the NS ($\sim 10^7 - 10^8$ years). \citet{pgw2006} and later \citet{vig2013} have also shown that around $\sim 10^6$ years, the surface temperature remains around $10^5$K. In binary X-ray pulsars, there are constant thermo-nuclear reactions taking place at the surface of the star due to in-falling accreted matter on the NS. Isolated radio pulsars do not have such external sources to hugely affect their thermal states. Said so, there are proposed models about the in-falling plasma of charged particles (the anti-particles of the ones that take part in forming the pulsar emission) on the polar caps and heating up the polar caps considerably (\citet{tsg2017}; \citet{szge2020}). Even if we take into account these effects, since the process is a continuous one, the entire system sets into a thermal equilibrium. 

Hence for a short time evolution and variability study of the surface magnetic field, isothermal state of the NS crust is a reasonable approximation. For our present analysis, we have taken a constant temperature of $10^{5.2}{\rm K}.$

\section{Results and Discussion}
\label{sec:res}

\begin{figure}
  \includegraphics[keepaspectratio=true,scale=0.33, angle=-90]{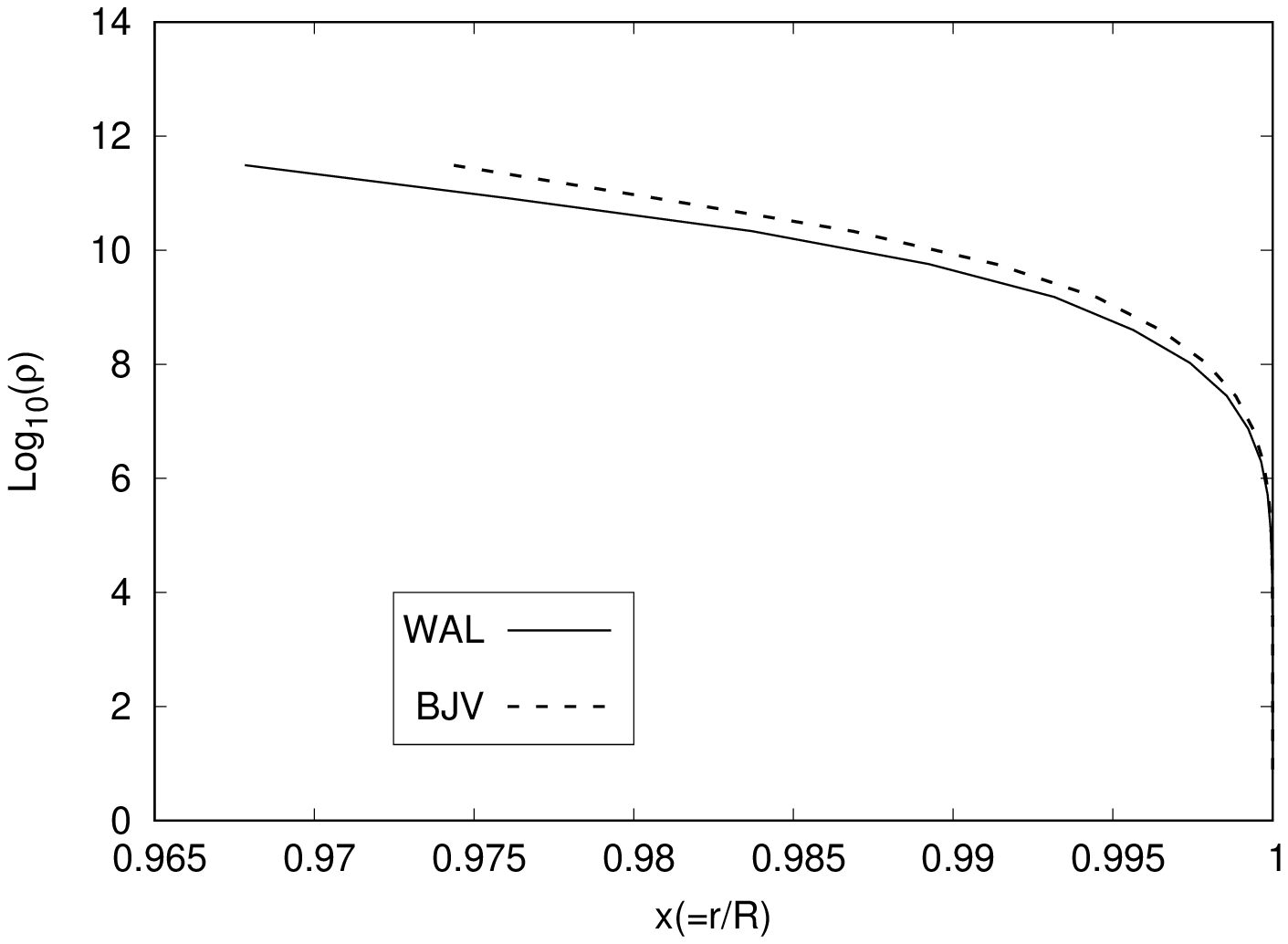}
\caption{Density profile of the outer crust for Walecka model (WAL) and Bethe-Johnson model V (BJV) (data taken from \citet{dattaetal1995}). }
\label{fig:dtb_rho_r}
\end{figure}

In our calculations, we initially considered two equation of states of matter - the Walecka model (tagged as `WAL' in this paper), which is a softer equation of state and the Bethe-Johnson model V (tagged as `BJV'), which is a relatively stiffer equation of state.\footnote{\cite{wal1974} formulated the equation of state for pure nuclear matter using scalar and vector meson interactions. On the other hand \cite{bjv1974} devised phenomenological potentials for neucleon-nucleon interactions to devise their equation of state.} The outer layer of the NS crust starts from the nuclear drip region - which is effectively from a depth of only a few hundred metres - corresponding to only 4\% of the stellar radius. \cite{dattaetal1995} carried out a numerical survey of the crustal density profiles for various EOSs. We used their tabulated data of the crustal density profile for the 1.4~$M_\odot$ stars, and plotted them in Fig.~\ref{fig:dtb_rho_r}.

\begin{figure}
  \includegraphics[keepaspectratio=true,scale=0.33, angle=-90]{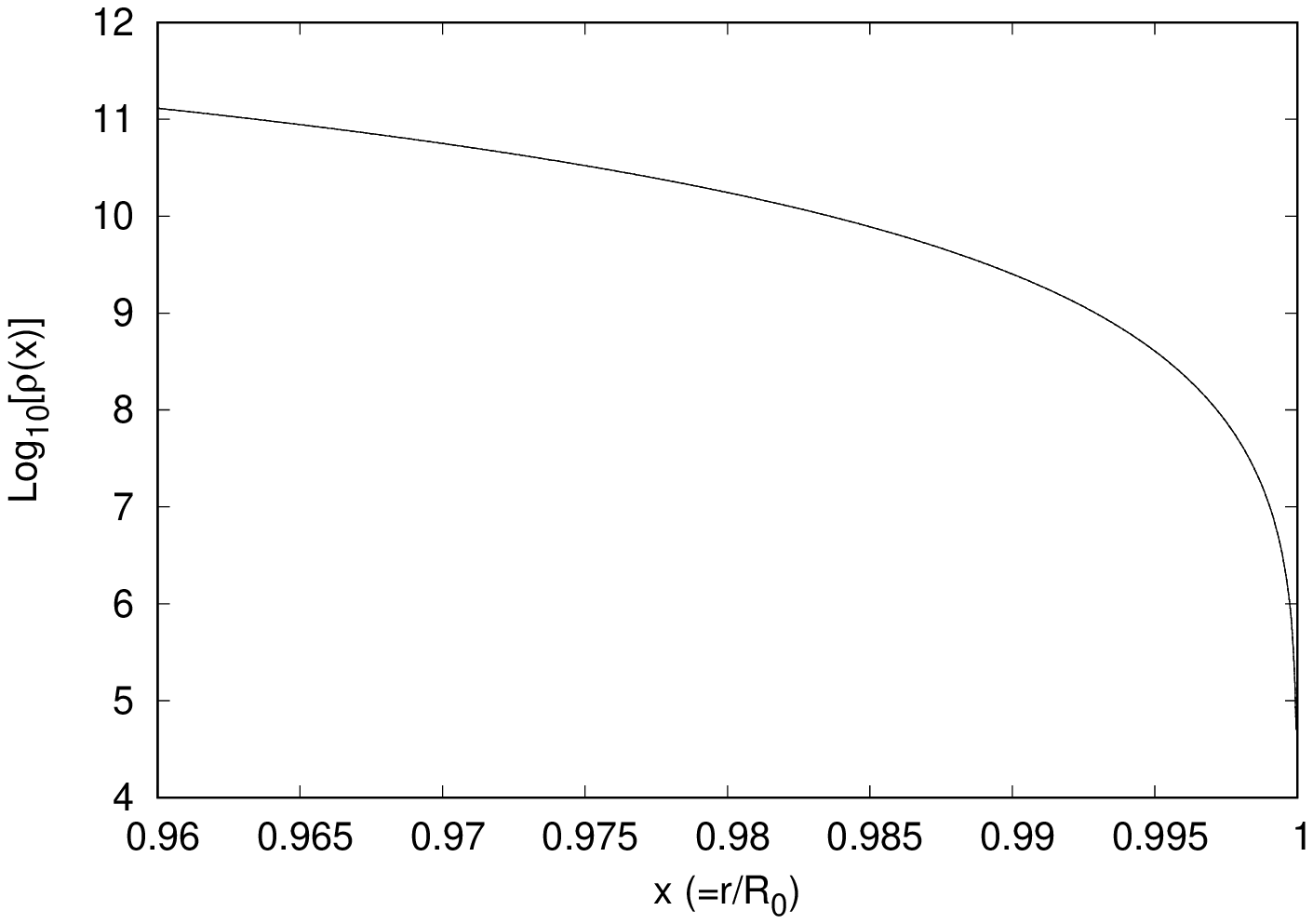}
\caption{Crustal density profile given by equation (\ref{eq:rhor}).}
\label{fig:den_vs_rho}
\end{figure}

The model of the density profile given in equation~(\ref{eq:rhor}) derived from \citet{uryak1979}  has been plotted in Fig.~\ref{fig:den_vs_rho}. The mass and radius parameters chosen in equation~(\ref{eq:rhor}) are 1.4~$M_\odot$ and 12.28~km respectively, representing values of the Walecka model (WAL) as tabulated in \cite{dattaetal1995}. The two figures show a close resemblance - this justifies the choice of the density profile obtained in equation~(\ref{eq:rhor}). In the subsequent results and plots, we will primarily use the NS parameters following WAL. 

\begin{table*}
\centering
\caption{Values of $\Delta B$ in Gauss for a $0.2\%$ fluctuation in the crust width for WAL given for different points in time with two values each of $Q$ and $T$ for the flat and curved spacetimes.}
\begin{tabular}{| c | c | c | c | c | c | c | c |}
\hline
\multicolumn{2}{|c|}{\bf Time [years]} & \multicolumn{2}{|c|}{$\mathbf{10^3}$} & \multicolumn{2}{|c|}{$\mathbf{10^5}$} & \multicolumn{2}{|c|}{$\mathbf{10^7}$} \\
\hline
 & \backslashbox{\bf Q}{\bf T [K]} & $\mathbf{10^{5.2}}$ & $\mathbf{10^{5.5}}$ & $\mathbf{10^{5.2}}$ & $\mathbf{10^{5.5}}$ & $\mathbf{10^{5.2}}$ & $\mathbf{10^{5.5}}$ \\
\hline
\multirow{2}{*}{\bf Flat} & $\mathbf{0.001}$ & $1.276\times 10^8 $ & $1.685 \times 10^8$ & $3.178\times 10^8  $ & $3.981\times 10^8  $ & $5.431\times 10^8  $ & $5.094\times 10^8  $ \\ \cline{2-8}
                                 & $\mathbf{0.1}$ & $1.511\times 10^8  $ & $1.831\times 10^8  $ & $4.652\times 10^8  $ & $4.863\times 10^8 $ & $4.6225\times 10^8 $ & $4.4595\times 10^8 $ \\
\hline
\multirow{2}{*}{\bf Curved} & $\mathbf{0.001}$ & $1.14\times 10^8 $ & $3.27\times 10^8 $ & $3.46\times 10^8 $ & $9.02\times 10^8 $ & $7.009\times 10^8 $ & $1.413\times 10^9 $ \\ \cline{2-8}
                                 & $\mathbf{0.1}$ & $2.73\times 10^8 $ & $3.55\times 10^8 $ & $1.05\times 10^9 $ & $1.127\times 10^9$ & $1.1616\times 10^9 $ & $1.1546\times 10^9 $ \\
\hline

\end{tabular}
\label{tab:Wal}
\end{table*}

We solved the differential equations (\ref{eq:flatdiff}) and (\ref{eq:curveddiff}) for the flat and curved spacetime background, in MatLab using the \emph{Method of Lines} based on finite differencing approximations of spacial derivatives. The resulting solutions for $f(x,t)$ and $g(x,t)$ gave us the corresponding vector potentials from which we calculated the corresponding surface magnetic fields. The surface magnetic field decay for both the flat and the curved spacetimes are found to be spanning over a period $10^{8}$ to $10^{9}$ years. In order to verify the consistency of our numerical results, we have compared them with the results of the surface magnetic field decay study as obtained by  \cite{batdat1996} (for non-GR cases) and \cite{sen1997} (including the GR effects). Here we also notice that the general relativistic effects delay the magnetic field decay time.  Essentially, the additional effects from general relativistic terms come mostly from the gravitational potential $(1-2m/r)$ that add up to the stellar compactness. When compared to the flat spacetime (Eq~(\ref{eq:flatdiff})), the diffusion equation for the curved spacetime (Eq~(\ref{eq:curveddiff})) is affected by the  softening of the l.h.s of the equation - thus allowing a larger value of time in the solution matrix.  Hence for a given period of time, the surface magnetic field decay calculated in a GR background is slower, thus extending the decay time.

We next consider the effects radial oscillations may have on the surface magnetic field. A simple form for the radial oscillatory motion can be  described by $$ \delta x (r,t) = X(r){\rm  exp}(i\omega t). $$
The frequency ($\omega$) of radial oscillations of a NS is primarily driven by the collective excitation of nucleons leading to the giant iso-scalar monopole oscillations, better known as the breathing mode of oscillations, in the inner core of the star. Hence the frequency of the radial oscillation is dependent on the EoS of matter in the core. This frequency is also affected by a small contribution from the gravitational forces. The amplitude of this oscillation $X(r)$ is however not so straight forward across the entire radius of the NS as it is expected to have a variable amplitude that is more prominent across the low density regime below the nuclear drip region of the outer crust. \cite{gll1983, vc1992} \& \cite{koru2001} calculated the frequencies of radial oscillation for different known nuclear matter EoSs where it has been shown that these frequencies vary larger than a kilohertz and more (periods of microsecond to nanosecond). Absence of any previous estimation of the amplitude variation across the entire radius of the NS (from literature) prompted us to consider a two fluid setup where the inner fluid is of the dense nuclear matter with high incompressibility, and the outer fluid is of the crustal matter with lower incompressibility. Then during radial oscillation of the NS, the outer crust should oscillate with a larger amplitude than the rigid inner core. A full consideration of a two fluid model would be of interest in order to estimate the amplitude of oscillations where the outer crust oscillates with a larger amplitude than the rigid inner core. However, for this work we consider only fluctuations in the outer crust (as this is where they will be most prominent) as a first step to see if this consideration is viable.

In the next step we study the variation of the surface magnetic field due to variations in the outer crustal radius of $±0.2\%$ and $±0.4\%$ (which translates to radial variations of only $1$ metre and $2$ metre respectively). In order to fluctuate only the width of the crust, the initial depth, $x_0$, is taken to be the same physical distance from the centre of the star for the perturbed fields $B_+$ and $B_-$ as it is for the unperturbed field $B_0$. So for the unperturbed field, the initial depth corresponds to a radius of $0.96 \times R$. The initial depth values for the perturbed fields $B_+$ and $B_-$ with total radii of $R_+$ and $R_-$ respectively are then given by $x_0 = 0.96 \times (R/R_+)$ and $x_0 = 0.96 \times (R/R_-)$ respectively. The total radii are given by the addition (or subtraction) of $0.2\%$ and $0.4\%$ of the crust width such that for a $0.2\%$ variation in crust width the radii are given by $R_+ = R + R \times 0.04 \times 0.002$ and $R_- = R - R \times 0.04 \times 0.002$, and for a $0.4\%$ variation in crust width the radii are given by $R_+ = R + R \times 0.04 \times 0.004$ and $R_- = R - R \times 0.04 \times 0.004$.

\begin{figure}
    \includegraphics[width=0.7\linewidth,angle=-90]{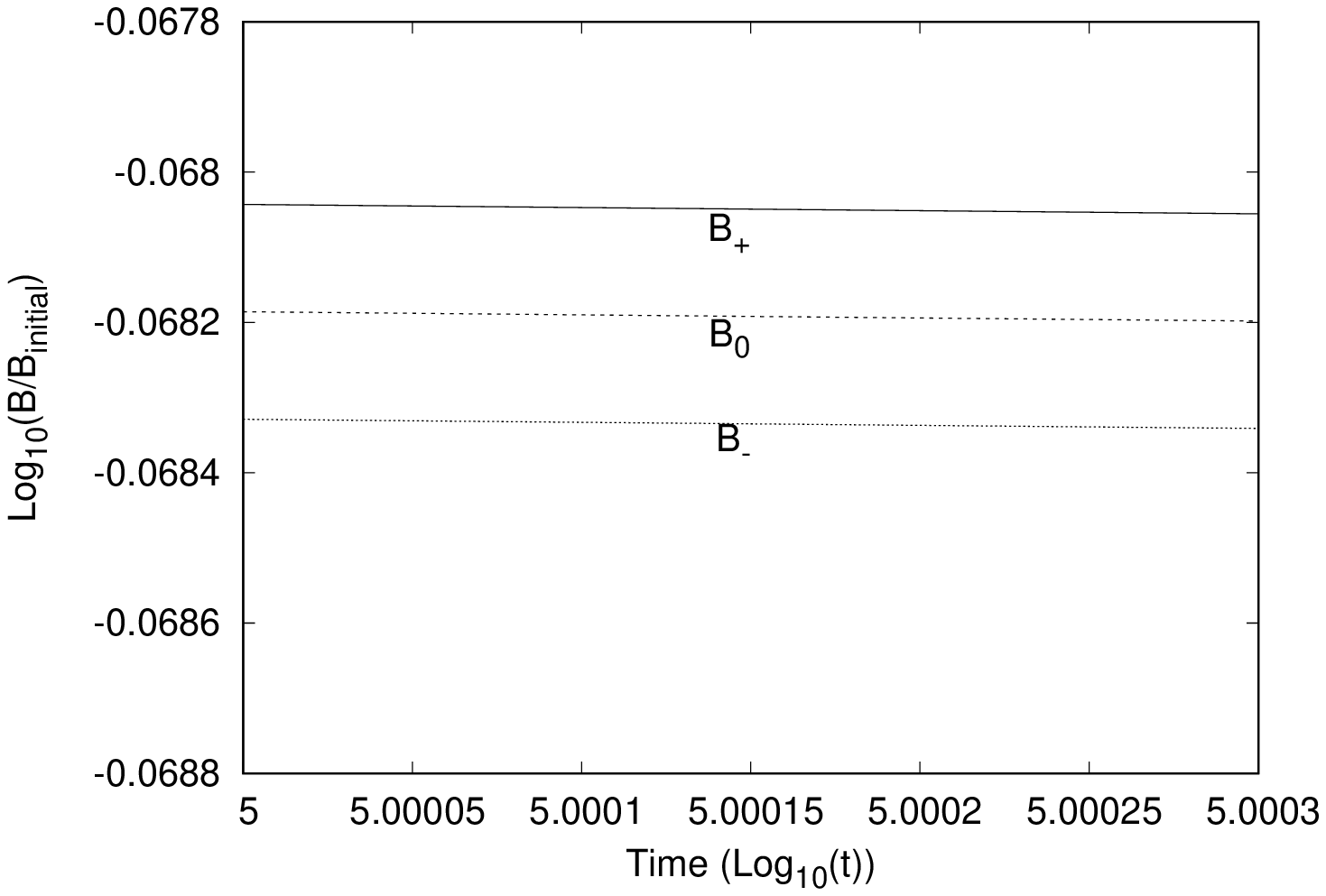}
\caption{Variation in the surface magnetic field due to the radial fluctuation of 0.2\%  and 0.4\% of the outer crust. $B_0$ is the non fluctuating field and $B_+$ and $B_-$ are the values for maximum and minimum radial variations respectively.}
\label{fig:B_rad_fluc}
\end{figure}

As the star oscillates radially the magnetic field strength will oscillate between the minimum and maximum values with a frequency equivalent to the radial oscillation frequency. In Figure~\ref{fig:B_rad_fluc} we have shown this resultant variation on the surface magnetic field for a $0.2\%$ fluctuation of the crustal width, in a time span of $\sim 70$ years for a NS aged at $10^5$ years in a curved spacetime background. The lines corresponding to $B_0$ are the surface magnetic field for the unperturbed radius, $B_+$ indicates the surface field for an expanded radius, and $B_-$ for a contracted one. All magnetic field values have been normalised by the initial value of the magnetic field $B_i$.

As expected, the expanded radius (and hence lowering of the density) reduces the conductivity, resulting in an increased surface magnetic field ($B_+$), and a contracted radius results in a decreased growth of the surface magnetic field ($B_-$).

\begin{figure}
  \includegraphics[keepaspectratio=true,scale=0.33, angle=-90]{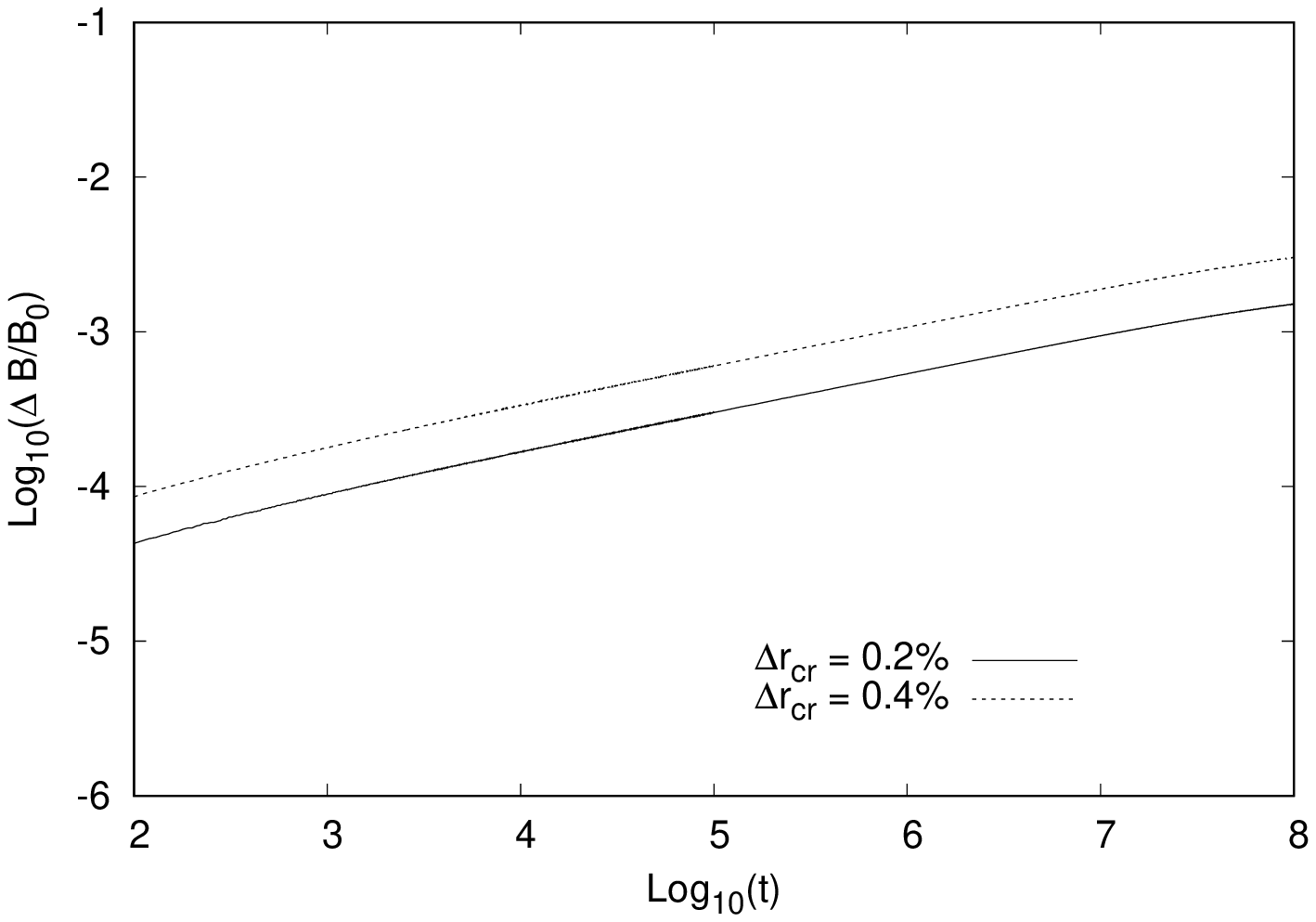}
\caption{Variation in the surface magnetic field $\Delta B$ due to the radial fluctuation of 0.2\%  and 0.4\% of the outer crust, including the general relativistic effects. Notable is the increase in the magnetic field fluctuation in the later stage of the NS's lifetime}
\label{fig:DelBbyB_cur}
\end{figure}

We have then studied how the change in the magnetic field ($\Delta B$) evolves with time over the lifetime of a NS and how this may be interpreted observationally. In Figure \ref{fig:DelBbyB_cur} we have plotted the variation of the surface magnetic field $\Delta B$ relative to the unperturbed value of the field $B_0$ due to $0.2\%$ and $0.4\%$ fluctuation of the crustal width, for the entire time period of $10^2$ to $10^8$ years of a thermally equilibrated NS's life.

Table \ref{tab:Wal} shows the values of $\Delta B$ in Gauss for a $0.2\%$ fluctuation in the crust width for WAL. The values are shown for three different ages in the NS's lifetime ($t = 10^3$ years, $t=10^5$ years and $t=10^7$ years) to demonstrate the magnitude of the change in the magnetic field at different stages in the evolution. This is done for the flat and curved spacetimes with two values each of $Q$ and $T$. It can be seen that this magnitude of $\Delta B$ is of the of the order of $\sim 10^8$ G $- 10^9$ G with values varying by an order of magnitude over the NS lifetime. The consequences of this change in $\Delta B$ will be further explored and made clearer in the following discussions.

Studies were done for both the flat and curved spacetime cases comparing the EOSs WAL and BJV and for $0.2\%$ and $0.4\%$ variation in the crust width. The numerous plots of which will be neglected her for the sake of brevity, however, the general findings can be noted. The fluctuations in the magnetic field are naturally related to the size of the radial fluctuations, but the compactness of the star also has an effect. The softer EoS of WAL has a sightly larger variation in magnitude of the magnetic field than BJV. For the flat spacetime this can simply be attributed to the fact that for a given mass, WAL has a larger radius than BJV and hence the fluctuations in the crust width are also large. However, the effect of curvature enhances the differences in the EoSs as the gravitational potential adds to the stellar compactness. The more compact the star, the smaller the fluctuations in the magnetic field. This is consistent with our assumption that the core of the star will fluctuate less than the crust due to its high density and compactness increasing the stiffness of the matter, and hence the majority of the fluctuations in the radius will occur in the outer crust.

As well as the compactness of the star having an effect on the magnitude of the oscillations, the magnitude of the unperturbed magnetic field also plays a significant role.  The magnetic field of a NS decays as the star ages. As the magnetic field decreases, we find (as seen in Figure~\ref{fig:DelBbyB_cur}) that the change in the surface magnetic field $\Delta B$ increases. Thus any microstructure caused by the radial oscillations will be more noticeable in the later stages of the star's evolution. This is consistent with observational data which shows that drifting sub-pulses are more prominent closer to the deathline \citep{song2023}.

\citet{song2023} conducted a study on 1198 pulsars using the Thousand-Pulsar-Array programme on MeerKAT. They found that 418 pulsars ($\sim 35\%$) of these pulsars exhibit drifting sub-pulses and that these sub-pulses are more pronounced towards the deathline, as is consistent with other previous studies \citep{ran1986,welt2006}. The detectability of drifting sub-pulses is dependant on the signal to noise ratio (S/N) which is higher for long period pulsars associated with larger characteristic age $\tau_c$. It is found that the younger pulsars produce more erratic or diffuse behaviour, whereas the older pulsars exhibit more coherent sub-pulse modulations. They also note that drifting sub-pulses are a very common occurrence and  if high quality data were available for all pulsars then this phenomena should be present in $\sim 60\%$ them. This suggests that the phenomena should be related to some common feature in NSs. 

The preference for the detection of drifting sub-pulses in older and less energetic pulsars suggests that drifting sub-pulses become stronger as the pulsars age. Since these drifting sub-pulses are so common it is interpreted that the physical processes behind the production of said sub-pulses should be related in some way to the radio emission mechanism and may well be a fundamental property of the emission mechanism. In many works, the drifting sub-pulses are explained by the geometric interpretations of the carousel model, however, they could be explained by any model that is able to produce such periodic sub-pulse modulations, such as the radial oscillations of the star. We suggest that the physical process behind these sub-pulses could be explained by the radial oscillations and the fact that the sub-pulses are pronounced toward the deathline can be explained by the fact that, from our study, it is seen that the older pulsars have a more pronounced change in the magnetic field strength due to the radial oscillations, thus any sub-pulses produced from this phenomena would be more pronounced as the pulsar ages.

In our analysis, we have considered that the magnetic field of the NS is located in the outer crust. Without imposing such stiff constraint, even if we consider the magnetic field is located in the deeper layers, due to the high electrical conductivity, the field strength there will be relatively weaker than that in the low density region of the outer crust. Hence, majority of the contribution to the surface magnetic field will come from the outer crust, and other than fine tuning the values for fields present in the deeper layers, qualitatively our results will remain the same.

\section{Conclusion}
\label{sec:concl}

We considered the effect of radial oscillations on the surface magnetic field of a NS and how this may be linked to observational phenomena. We found the fluctuations of the magnetic field strength due to these oscillations on a short timescale to be of a significant order such that we may be able to observe these fluctuations in the form of microstructure and sub-pulses in a pulsars observed emission. In the long term, we found the change in the magnitude of these fluctuations in the magnetic field to increase with time and thus any sub-pulses caused by these fluctuations should be more noticeable as a star ages. In other words older stars should be more effected by the radial oscillations and would exhibit more pronounced microstructure and sub-pulse modulations than younger stars. It is exciting to find that our results agrees with observational data as is shown in the very recent work of \citet{song2023} who found that drifting sub-pulses are more pronounced in older pulsars towards the deathline. We may also note that the intensity of the varying surface field is directly related to size of the fluctuation of the radial length of the crust as can be seen from the $±0.2\%$ and $±0.4\%$ variations in crust width. Additionally to this, the GR effects add to the compactness of the star which leads to smaller fluctuations in the magnetic field strength.

\section*{acknowledgements} 

KS acknowledges S Shindin of UKZN for helping with the computational codes. KS also acknowledges the NRF for partial research support. SR thanks D Mitra of NCRA, Pune, for many insightful discussions. 

\section*{Data Availability}
No new data were generated or analysed in support of this research.

%

\bsp	
\label{lastpage}
\end{document}